# Highly Conductive RuO$_2$ Thin Films from Novel Facile Aqueous Chemical Solution Deposition

Martina Angermann[1], Georg Jakopic[2], Christine Prietl[2], Thomas Griesser[3], Klaus Reichmann[4], Marco Deluca[1]

[1] *Materials Center Leoben Forschung GmbH, Department of Microelectronics, Leoben, Austria (corresponding author: martina.angermann@mcl.at)*

[2] *Joanneum Research Forschungsgesellschaft mbH, Institute for Sensors, Photonics and Manufacturing Technologies*

[3] *Montanuniversität Leoben, Institute of Chemistry of Polymeric Materials*

[4] *Graz University of Technology, Institute for Chemistry and Technology of Materials*

## Abstract

Ruthenium dioxide (RuO$_2$) thin films were synthesized by Chemical Solution Deposition (CSD) on silicon substrates using only water and acetic acid as solvents. The microstructure, phase-purity, electrical and optical properties as well as the thermal stability of the thin films have been characterized. The microstructure of the thin films strongly depends on the annealing temperature: A smooth thin film was achieved at an annealing temperature of 600°C. Higher annealing temperatures (800°C) led to radial grain growth and an inhomogeneous thin film. A very low resistivity of 0.89 μΩm was measured for a 220 nm-thick thin film prepared at 600°. The resistivity of the thin films increases with temperature, which indicates metallic behavior. Phase-purity of the thin films was confirmed with X-ray Diffraction (XRD) measurements, X-ray Photoelectron Spectroscopy (XPS) and Raman spectroscopy. Transmission and reflectivity measurements indicate that RuO$_2$ efficiently blocks the UV-VIS and IR wavelengths. The optical constants determined via spectroscopic ellipsometry show high absorption in the near-IR region as well as a lower one in the UV-VIS region. The thermal stability was investigated by post-annealing, confirming that the thin films are stable up to 750°C in synthetic air.

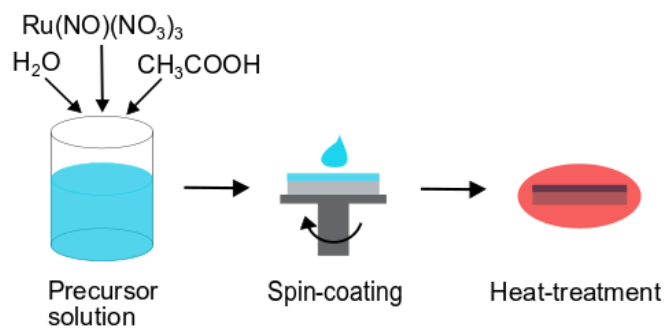

**Keywords**: ruthenium dioxide, chemical solution deposition, thin film, conductive metal oxides;

**Acknowledgements:** This project has received funding from the European Union's Horizon 2020 research and innovation programme under grant agreement No 951774. The authors want to thank J. S. Mateo (XRD measurements of the powders), K. Bakken, T. Gindel, A. Kobald and H. Kobald of the Materials Center Leoben Forschung GmbH for their collaboration and fruitful discussions.

**Highlights**

- RuO$_2$ thin films were prepared via a facile CSD method using simply water and acetic acid as solvents, avoiding more hazardous solvents and additives.
- Raman and XRD spectra of the thin films confirm their phase-purity.
- SEM analysis shows that the microstructure can be tuned with annealing temperature.
- The thin films have a very low specific resistivity of 0.89 µΩm.
- Electrical measurements suggest that the thin film is stable in air up to 750°C for 30 min.
- Optical properties show a high absorption in the NIR-region tunable with film thickness as well as a lower absorption in the UV-VIS region.

# 1 Introduction

Conductive metal oxides are interesting for a wide range of applications such as ferroelectric and magnetoresistive devices, solar cells, solid oxide fuel cells (SOFC) and sensors, owing to their excellent chemical and thermal stability. Among the metal oxides, ruthenium dioxide (RuO$_2$) is particularly interesting since it offers high conductivity (i.e. low specific resistivity, 0.35 µΩm, at room temperature, RT, single crystal, [1]) and chemical stability. Hence, it is an important material as an electrode for (electro)catalysis (e.g. oxidation reactions, electrolysis,..), energy storage (supercapacitors, Li-ion batteries,..) and semiconductor devices (memristors, gate contacts, [2]). In capacitor structures RuO$_2$ electrodes moreover reduce leakage currents, resulting in negligible fatigue and low retention losses, compared to metal electrodes, such as Pt [3–6]. Due to their wide range of applications, RuO$_2$ thin films have been prepared by a number of techniques such as metal organic vapor deposition (MOCVD) [7], sputtering [8]–[11], pulsed laser deposition (PLD) [12], atomic laser deposition (ALD) [13],[14] and electrodeposition [15]. Sol-gel deposition of thin films is cheaper and more flexible compared to the aforementioned physical approaches. However, despite the apparent advantages, there are currently only few reports on the sol-gel preparation of RuO$_2$ thin films: In 1996 Watanabe et al. prepared RuO$_2$ thin films by dissolving ruthenium chloride n-hydrate in ethanol and depositing by spin-coating on Si substrates with different annealing temperatures. They obtained a specific resistivity of 1.8 µΩm by adjusting annealing temperature to 800°C, which led to grain sizes of ~200 nm [16]. In 1997 Teowee et al. prepared 1 µm thick RuO$_2$ films with 5 µΩm on silicon substrates, using ruthenium chloride n-hydrate in ethanol [17]. 1998 Yi et al. used Ru nitrosyl nitrate and 2-methoxyethanol to deposit thin films on silicon substrates. They lowered the resistivity of the thin films down to 2 µΩm by increasing the thickness to 450 nm and the low resistivity was almost constant for annealing temperatures between 500-700°C [18]. In the following year, Yi et al. used the same precursor solution to deposit thin films on stainless steel and achieved lower resistivities with even thinner films (240 nm) [4]. In 2002 Armelao et al. used dip-coating from alcoholic solutions of Ru(OEt)$_3$ to prepare thin films and analyzed the thin films using XRD and XPS [19]. In 1999 Bhaskar et al. used the same components as Teowee et al. to prepare 700 nm thick films with 2.94 µΩm by adjusting the annealing temperature to 700°C [20]. To the best of our knowledge, no sol-gel deposition using non-toxic aqueous solutions for the preparation of RuO$_2$ thin films was reported so far. Such a safe and low-cost approach offers great advantages for industrial applications.

In this work, we present a novel simple approach to deposit RuO$_2$ thin films using only the environmentally-benign solvents water and acetic acid. The microstructure, electrical, thermal and optical properties of the thin films were also studied and the thin film obtained showed excellent specific resistivity values of 0.89 µΩm with a thickness of 220 nm.

## 2 Experimental

For RuO$_2$ thin film deposition, a 0.4 M solution was prepared by dissolving ruthenium(III)-nitrosylnitrate powder (Alfa Aesar, USA) in a 1:2 (V:V) water and acetic acid (Roth, Germany) mixture and stirring it overnight. The solution was deposited on plasma-cleaned silicon substrates (Si/600 nm <100>SiO$_2$, Siegert Wafer, Germany), which were subsequently spin-coated at 5000 rpm (with 2500 rpm/s rate) for 30 s. The thin films were dried at 160 °C for 5 min on a hotplate prior to heating to 350 °C (1°C/s, 2 min) and crystallized at higher temperatures (600/700/800 °C, 10°C/s, 10 min) in a rapid thermal annealer (MILA-5050, ULVAC GmbH, Germany) under a constant gas flow of 0.8 l/min of N$_2$ and 0.2 l/min of O$_2$, corresponding to synthetic air. The deposition cycle was repeated 10 times to yield a thickness of ~200 nm. Additional post-annealing in the rapid thermal annealer was done for some of the samples. For the optical characterization also samples on fused silica substrates (MicroChemicals, Germany) were prepared with the same procedure. RuO$_2$ powder was prepared by annealing the dried solution at 600/700/800/900 °C for 2h in a muffle oven and crushing the powder in an agate mortar.

The thermal behavior of dried gel and powder was characterized with a TGA-DSC-MS (STA449F1A, coupled to a QMS 403c mass spectrometer, Netzsch, Germany) using a heating rate of 5°C/min. Raman measurements were performed with a WITec alpha300R spectrometer (WITec GmbH, Ulm, Germany) with 1800 gr/mm and an EC Epiplan-Neofluar DIC objective (Zeiss, Germany) using 10 mW of a 532 nm laser. Powder XRD was performed with the D2 Phaser (Bruker, Germany) using a Co-K$_\alpha$ source with 0.06° per step between 15° and 90°. For the thin films Cu-K$_\alpha$ source grazing incidence XRD (GI-XRD) was done with the D8-Discover Series II (Bruker, Germany) with 0.03° per step between 10° and 80°. X-ray photoelectron spectroscopy (XPS) was conducted with a Thermo Fisher Scientific Inc. Nexsa G2 photoelectron spectrometer system equipped with a low-power Al-K$_\alpha$ X-ray source yielding a 30-400 µm adjustable X-ray spot size. Scanning Electron Microscope (SEM) images were recorded with the Auriga 40 (Zeiss, Germany). The transmission and reflectance of RuO$_2$ thin films in the UV to near infrared range were measured with a Lambda 900 spectrometer (Perkin Elmer, Great Britain). Infrared spectra were measured up to 16 µm using a FTIR Bruker Tensor 27 Instrument. Spectroscopic ellipsometry was used to determine the optical constants of the thin films (instrument J.A.Woollam VASE with the proprietary software). The spectral range extended from 300 nm to 1700 nm with a step size of 5 nm, and 65°, 70°, and 75° were used as the angles of incidence for the measurement. A homogeneous layer model was applied to evaluate the measurements, the dielectric function of which consisted of a DC offset, 5 Gaussian-broadened oscillators, and a pole point in the far-infrared. With this model, the measured values could be reproduced very well. Moreover, the sheet resistance was measured with the 4-point-probe setup of the aixACCT TF Analyzer 3000 (aixACCT Systems GmbH, Aachen, Germany) using tungsten needles. The thickness needed for the calculation of resistivity was determined from SEM images of cleaved samples.

## 3 Results and Discussion
### 3.1 Thermal Analysis

For the DSC-TGA measurements the solution was dried at 200°C or calcined at 600°C in a muffle furnace for 2 h, both in air. For the powder prepared at 200°C there is an exothermic peak visible around 280°C, which is likely due to the pyrolysis of the material (see Fig. 1). Also, a drastic mass loss of 45% is recorded between 170°C-320°C, which could be due to gas evolution (e.g. CO$_2$, NO$_x$, …). Such enormous gas evolution is typical for pyrolysis reactions. Theoretically, the mass loss of the conversion of ruthenium-nitrosylnitrate to RuO$_2$ should be 58%, which is relatively similar to the measured value. Measurement of the powder prepared at 600°C also exhibited an exothermic peak at 750°C (see Fig. 2), which was accompanied by a mass gain of ~18%. This could be linked to the oxidation of Ru-metal, which was present solely in the powder (see Fig. 3 in chapter 3.2). The kinetics of the conversion reaction in the thin films seem to be different to the prepared powder, since there was no Ru-metal visible in the XRD of the thin films. This indicates that the metal could be oxidized in air atmosphere to $RuO_2$, which is further confirmed by the fact that during the measurement in nitrogen atmosphere no mass change was observed. The fact that Ru metal was only recorded for the powders and not the thin films might be due to the different microstructure of both materials. Since each heat-treated thin film layer was only around 20 nm thick, oxygen probably could penetrate here the whole layer. However, in the case of the powders, $RuO_2$ might have first formed on the surface of the relatively large powder particles, and then acted as a diffusion barrier [21-22], which might have resulted in Ru metal being present in the core of the particles.

The information gained from the TGA/DSC measurements was used to design the temperature program for thin film preparation. Slow heating with a heating rate of 1°C/s was applied to the thin films up to the pyrolysis temperature of 350°C in order to avoid rapid gas evolution, which could lead to the formation of pores.

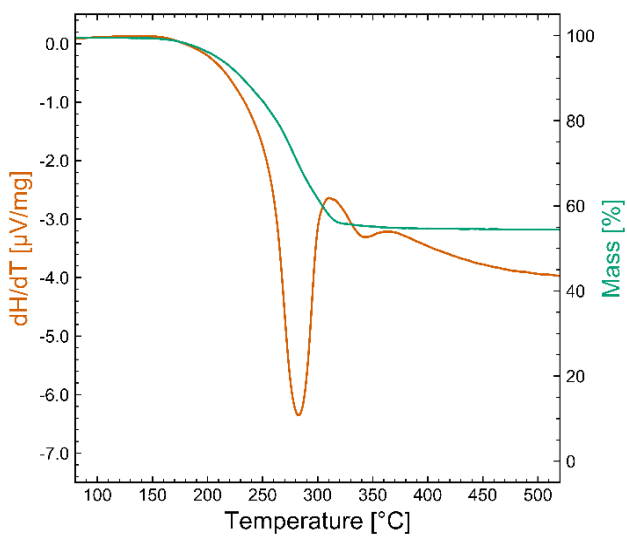

*Fig. 1* DSC-TGA measurements of the dried solution (exo ↓)

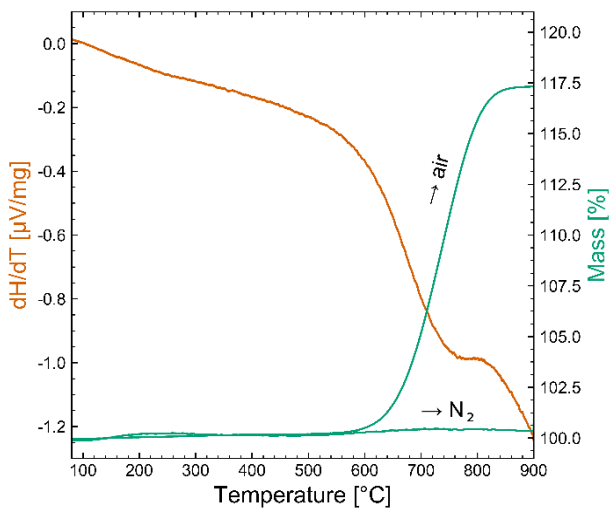

*Fig. 2* DSC-TGA measurements of the calcined (600°C) powder (exo ↓)

Additional TGA measurements coupled with mass spectrometry were done to identify the generated gases, and typical pyrolysis gases were detected (e.g. $CO_2$, $NO_x$, ...), with a high share of $CO_2$ due to the high amount of acetic acid in the solution (see supplementary information).

## 3.2 Phase Analysis

The phase purity and crystal structure of the thin films and powders was analyzed using XRD and Raman spectroscopy. Fig. 3 shows the XRD patterns of the powders prepared at different annealing temperatures, namely 600°C, 700°C, 800°C and 900°C. At annealing temperatures of 700°C or higher, peaks related to the planes (110), (101), (200), (111), (211),(110),(002), (221), (112), (301), (202) appear and the XRD-pattern matches well the reference spectra of $RuO_2$ with tetragonal rutile structure (COD Card 2101852 [23]), which confirms phase purity in the produced films. For powders annealed at the lower temperature of 600°C, additional peaks appear for 2Θ angles between 40° and 60°, which can be linked to Ru metal in the powder. The XRD-spectra also indicate that the crystallinity of the samples improves by increasing the annealing temperature.

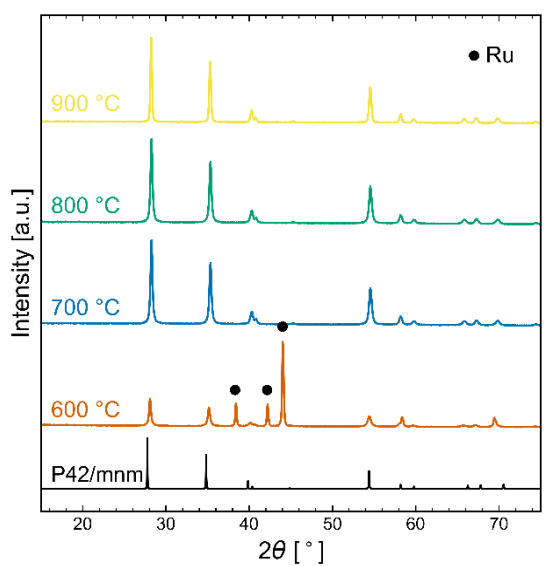

*Fig. 3* Normalized XRD patterns of the powders prepared at different annealing temperatures (600°C, 700°C, 800°C, 900°C) and rutile $RuO_2$ reference spectra (COD Card 2101852 [23]). The pattern has been converted to fit the Cu-$K_α$ reference, since a Cobalt-$K_α$ source has been used for measuring. Plotted with an offset for better visualization

GI-XRD was done on the thin films prepared with different annealing temperatures (see Fig. 4). The spectra show that phase pure and crystalline rutile $RuO_2$ thin films can be achieved even at relatively low annealing temperatures (600°C), which is confirmed by the sharpness of the peaks and the absence of Ru-metal peaks. The XRD pattern also indicates that there is a high (110) orientation, since the peak at ~28° is dominating.

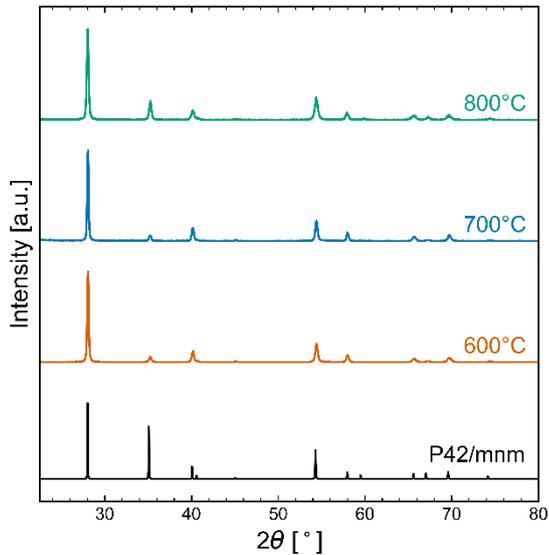

*Fig. 4* Normalized GI-XRD patterns of thin films deposited on silicon substrates annealed at 600°C, 700°C, 800°C. Rutile RuO$_2$ XRD patterns are given as reference (COD Card 2101852 [23], black curve). Plotted with an offset for better visualization

The Raman spectra of RuO$_2$ powders prepared at different temperatures are displayed in Fig. 5. The peaks can be assigned to the three major modes, $E_g$, $A_{1g}$ and $B_{2g}$, which are located at 528, 646 and 716 cm$^{-1}$ (cf. single crystal [24], [25]), respectively. The Raman spectra of powders calcined at 600°C were similar to the other ones, which shows that the Ru metal did not interfere with the measurement (as expected). The sharp peaks again indicate the high crystallinity of the samples that is present even at lower annealing temperatures. The weak first harmonics of $E_g$ and $A_{1g}$ modes can be also seen in the Raman spectra at 1016 cm$^{-1}$ and 1236 cm$^{-1}$ [26], respectively. There is a significant red shift of the peak positions of the three first-order Raman peaks. The shift increases with calcination temperatures, which might be due to increased strain states in the powder induced by the higher temperatures.

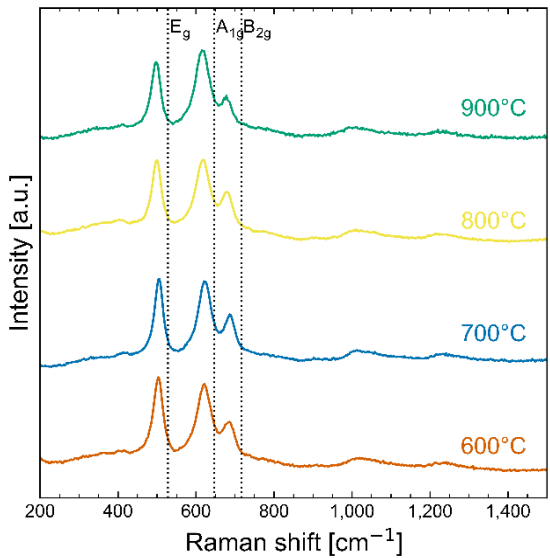

*Fig. 5* Normalized Raman spectra of the powders prepared at different temperatures (600°C, 700°C, 800°C, 900°C). The dashed lines indicate the 3 major Raman modes, *$E_g$, $A_{1g}$ and $B_{2g}$*. Plotted with an offset for better visualization

Also, for the thin films (cf. Fig. 6), a significant red shift of the first-order Raman modes was detected. This shift is likely to be attributed to a structural change in the RuO$_2$ lattice (similar to the powder case) and should not be related to any thermal mismatch to the substrate, since the volume thermal expansion coefficient of silicon (13.2·10$^{-6}$ °C$^{-1}$), is quite similar to that of RuO$_2$ 22.7·10$^{-6}$ °C$^{-1}$ [24].

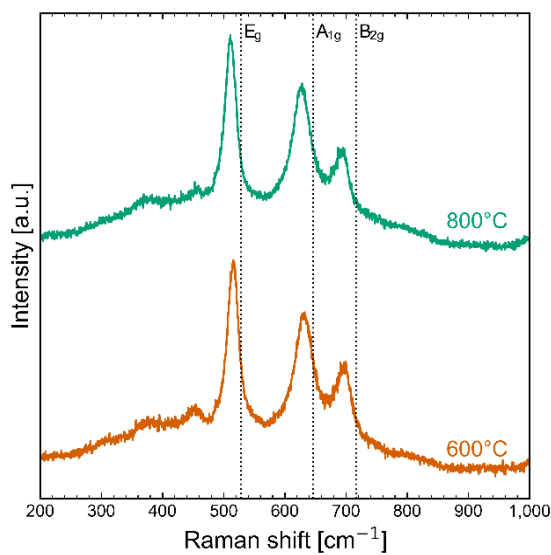

*Fig. 6* Normalized Raman spectra of the thin films prepared at different temperatures (600°C, 800°C). The dashed lines indicate the 3 major Raman modes, $E_g$, $A_{1g}$ and $B_{2g}$. Plotted with an offset for better visualization

Additionally, a XPS measurement of the thin film annealed at 600°C has been done to analyze the surface states of the thin film (see Fig. 7). The peaks of the convoluted fit of the measured data were assigned to specific photoelectrons (see Table 1). From the convoluted fit it can be concluded that the surface is phase-pure, and that the values are in good agreement to literature (doublet separation of 4.2 [27]). There are so-called satellite peaks visible in the XPS, which are often mis-assigned in literature to higher order oxides (RuO$_x$); however, as discussed by Morgan [27] these peaks are a result of spin-orbit coupling of non s-levels from the photoemission process, leading to this so-called satellite structure. A minor surface pollution with carbon was also detected (<6%), which is common in XPS investigations. No additional contaminants from the wafer (Si) or the precursor (Ru-nitrosyl-nitrate) were visible in the XPS, which shows that the CSD process is successfully creating a phase-pure thin film.

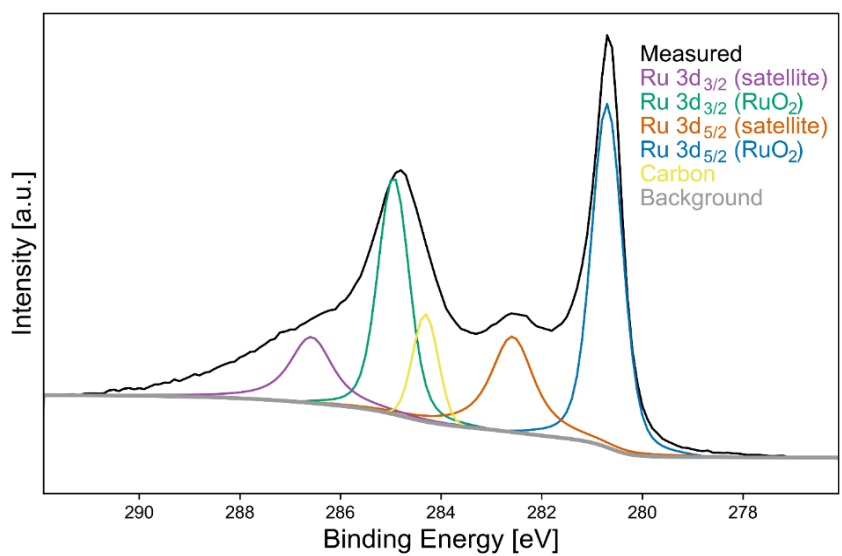

*Fig. 7* XPS spectra of a thin film prepared at 600°C (black) and convoluted assigned peaks of the fit

*Table 1* Overview of the compounds, binding energies and photoelectron orbital derived from the XPS measurements

| Compound | Binding Energy [eV] | Orbital |
|---|---|---|
| $RuO_2$ | 280.7 | $3d_{5/2}$ |
|  | 284.9 | $3d_{3/2}$ |
|  | 282.6 | $3d_{5/2}$ satellite |
|  | 286.6 | $3d_{3/2}$ satellite |
| Carbon | 284.3 | 1s |

In summary, the XRD, Raman and XPS measurements of the thin films suggest that the material is phase-pure and highly crystalline.

### 3.3 Microstructure

Scanning electron microscope (SEM) images of cross-sections of the thin films have been taken for different annealing temperatures (see Fig. 8). The microstructure looks very different, which indicates that the annealing temperature has a huge impact on the grain growth of $RuO_2$. The film heated to 600°C has a dense microstructure with columnar grains, the film heated to 700°C looks similar, but more grain boundaries are visible and the thin film heated up to 800°C displayed large round grains, which probably grew in radial direction at the expense of neighboring grains (i.e. Ostwald ripening). Consequently, the film heated to 800°C has a very rough surface. The microstructure also impacted the resistivities of the thin films: The dense and smooth film annealed at 600°C showed the lowest resistivity (see chapter 3.4). The total thickness could be obtained from the cross-section SEM images and was 220, 170 and 200 nm for the thin films prepared at 600, 700 and 800°C, respectively. However, due to the roughness of the thin film annealed at 800°C, the estimation of the thickness is prone to larger errors. The decrease in film thickness by increasing the annealing temperature from 600°C to 700°C could be due to a higher density from the higher ion mobility at increased temperature. A closer look at the microstructure of the different thin films also shows that there are smaller grains accumulated near the interface to the substrate. Hence, the first deposited layer might serve as seed layer, which promotes the growth of columnar grains from heterogeneous nucleation.

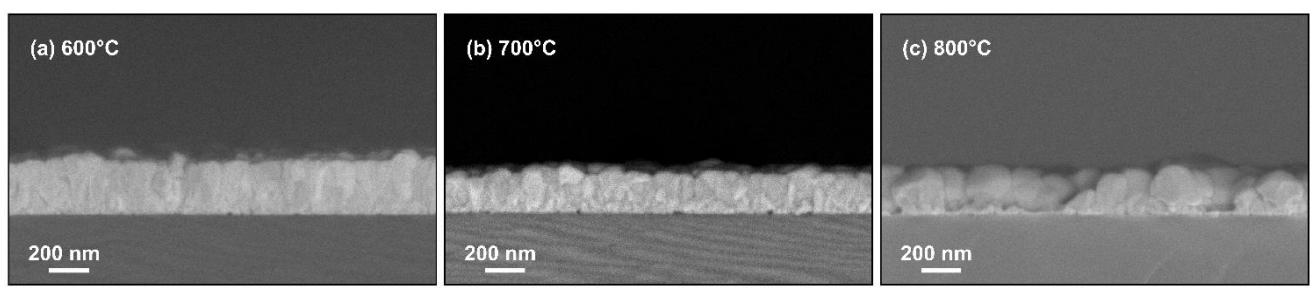

*Fig.* 8 SEM images of cross-sections of the thin films annealed at 600°C (a), 700°C (b) and 800°C (c) on top of the silicon substrate

Additional SEM images have been taken to check the influence of the heating rate and amount of acetic acid in the solution (see supplementary information), suggesting that the heating rate does not impact the microstructure significantly, and that a higher amount of acetic acid in the solution leads to a smoother microstructure with well-aligned columnar grains.

### 3.4 Electrical Properties

The resistivity of thin films is calculated using the thickness (t) of the films:

$$\rho = R \frac{\pi}{\ln(2)} t f_1 f_2 , \qquad (1)$$

Where R is the measured resistivity, $F_1$ and $f_2$ are geometric correction factors for non-negligible finite thickness compared to probe spacing, and finite sample dimensions to probe spacing, respectively.

Not only the microstructure, but also the resistivity of the thin films is highly influenced by the annealing temperature (see Table 2). Increasing the annealing temperature from 600°C to 800°C led to a 2.5 times higher specific resistivity. This is in accordance with the change in microstructure as already discussed in the previous section. The best resistivity was reached for 600°C and was 0.89 μΩm, which is much lower than previous deposition attempts using CSD but employing toxic 2-methoxyethanol as solvent [18]. The resistivity was measured three times by repositioning the needles, and its low error values indicate that the low resistivity of the thin films is not just a local phenomenon.

*Table 2* Specific resistivity of the RuO$_2$ thin films annealed at different temperatures. All thin films showed linear ohmic behavior during measurements. Each thin film consisted of 10 layers, and the total thickness was estimated via SEM images (600°C: 220 nm, 700°C: 170 nm, 800°C: 200 nm) and used to calculate the resistivity. The thickness of the sample from literature was 150 nm [18]

| Annealing Temperature [°C] | Specific Resistivity [μΩm] |
|---|---|
| 600 | 0.89 ± 0.06 |
| 700 | 1.03 ± 0.20 |
| 800 | 2.26 ± 0.25 |
| Literature | 2.7 [18] |

In Fig. 9 the specific resistivity and sheet resistance over thickness of the thin films are depicted for two annealing temperatures (600°C and 800°). The superior quality of the films annealed at 600°C is evident. Further, it can be clearly seen that the sheet resistance decreases with increasing thin film thickness in a similar manner for two annealing temperatures. From eq. (1) it is evident that the specific resistivity tends to saturate with increasing thickness, if the measured resistivity (R) is not decreasing significantly. Considering the results in Fig. 9, it can thus be concluded that the quality of the thin films remains good even after repeated depositions (>10 cycles).

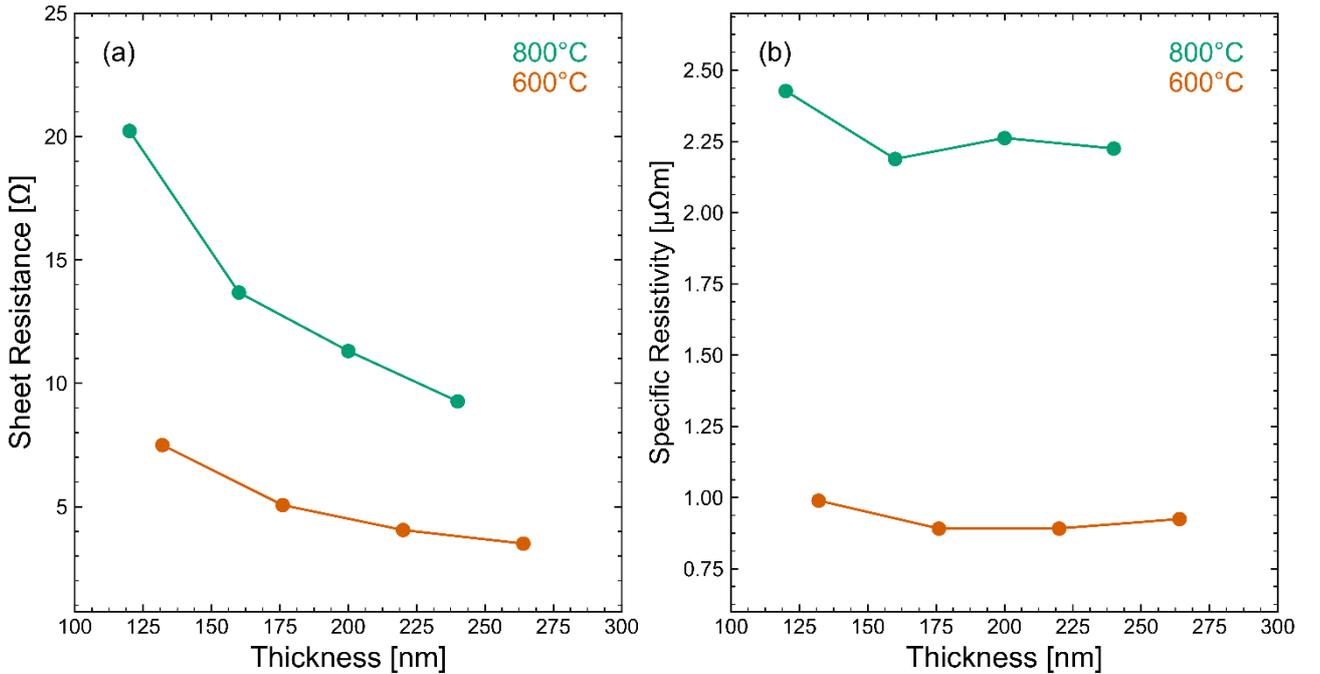

*Fig. 9* Sheet resistance (a) and resistivity (b) of thin films with increasing thickness. The same sample was measured between the repeated CSD steps, hence with increasing thickness

The temperature stability of a thin film prepared at 600°C was tested by measuring the resistivity and heating the sample to the desired temperature (see. Fig. 10). The RuO$_2$ thin film resistivity rises with temperature, which indicates metallic behavior. The linear fit (R$^2$=0.92) was used to calculate the temperature coefficient of resistivity, with the following formula:

$$RTC = \frac{\Delta \rho}{\rho_0 \Delta T}, \qquad (2)$$

where ρ is the resistivity, ρ$_0$ is the initial resistivity value and T is the temperature. The RTC value calculated is 5.8·10$^{-3}$ K$^{-1}$. This value is relatively high compared to literature values of RuO$_2$ thin films (~3 · 10$^{-3}$ K$^{-1}$ [28]) and other metallic thin films (200 nm thick Au, Cu, Al thin films: 3.36, 3.86, 3.86 · 10$^{-3}$ K$^{-1}$, respectively [29]), hence, the RuO$_2$ thin films might be interesting as thermistor material, for applications like temperature compensation circuits. After the heating cycle, the material was still highly conductive, since the resistivity reverted to the initial value. Hence, the metal oxide is highly stable, and can be used also for applications were stability against heat is necessary.

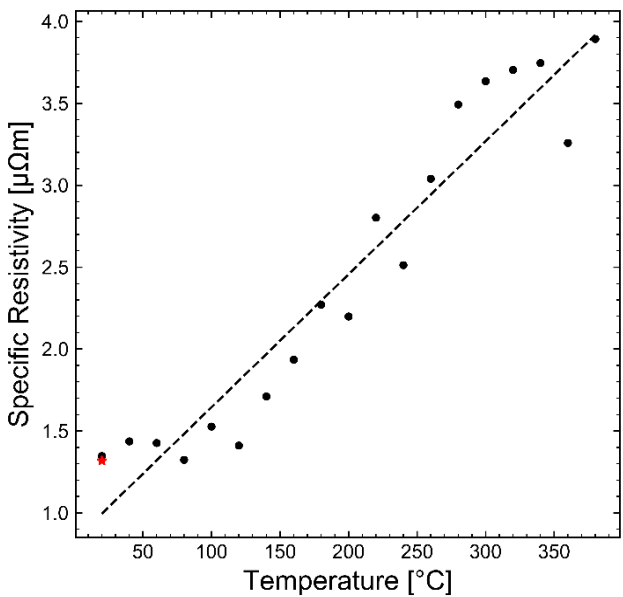

*Fig. 10* Change in resistivity with increasing temperature of a RuO$_2$ thin film prepared at 600°C. For every measurement the temperature was increased. The red star indicates the last measurement done after the heating cycle

### 3.5 High-Temperature Stability

The stability of RuO$_2$ at high temperatures (> 500°C) could be interesting for applications such as solid oxide fuel cells (SOFC), (chemical) sensors, micro electro-mechanical systems (MEMS) and for the use in harsh environments (e.g. geothermal applications, aerospace power electronics). According to literature, temperature stability is limited by the formation of gaseous RuO$_x$ in the presence of oxygen gases at temperatures above 800°C [30]–[32]. Hence, we investigated the resistivity of the thin films after post-annealing them at different temperatures: Post-annealing at 850°C in synthetic air for 1 h led to an increase in resistivity from 0.92 µΩm to 1.37 µΩm. The microstructure of the sample was investigated before and after the post-annealing and can be seen in Fig. 11. After annealing, the microstructure is much rougher, has additional pores and is thinned down in some areas. Moreover, an interface layer is visible in the SEM images, possibly due to interdiffusion. Repeating the post-annealing with oxygen atmosphere led to a discontinuous thin film with resistivities too high to be measured by the 4-point-probe method. These changes are likely caused by the oxidation of the RuO$_2$ with the formation of RuO$_x$ gases. In comparison, post-annealing in synthetic air at 750°C led to an unchanged resistivity, which shows that the thin film is stable in air even at such elevated temperatures.

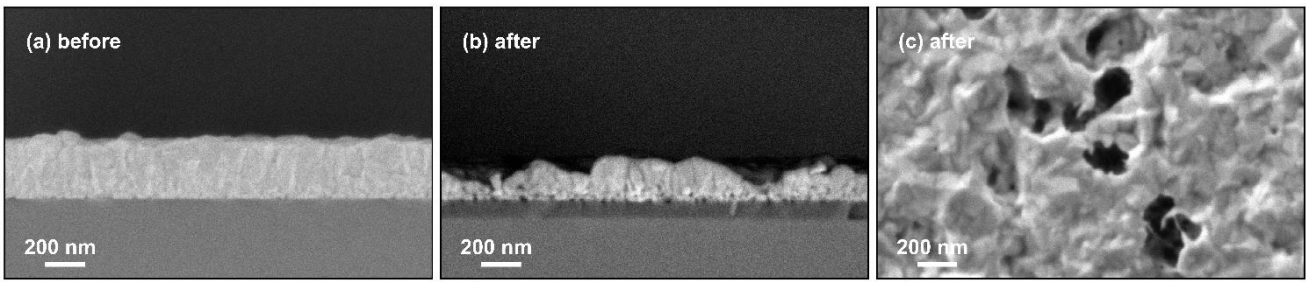

*Fig. 11* SEM images of the cross-sections of the thin film annealed at 600°C (a) and additionally post-annealed for 1h at 850°C in synthetic air (b) on top of the silicon substrate. (c) SEM image of the top view of the post-annealed thin film

## 3.6 Optical Properties

UV-VIS reflectivity and transmissivity spectra of $RuO_2$ thin films deposited on fused silica substrates with two different thicknesses – 22 nm (1 layer) and 220 nm (10 layers) - are displayed in Fig. 12 and Fig. 13, respectively. It can be seen that especially a 'thicker' layer of $RuO_2$ absorbs visible light well, since the reflectivity and transmissivity values are low in this wavelength range (A=1-R+T). This is in accordance with the observation that the thin films turned dark with increasing layers. The metallic character of the thin films detected in the electrical measurements is also indicated by the optical properties: There is high visible light absorption due to available energy states and surface electrons, which is typical for metals.

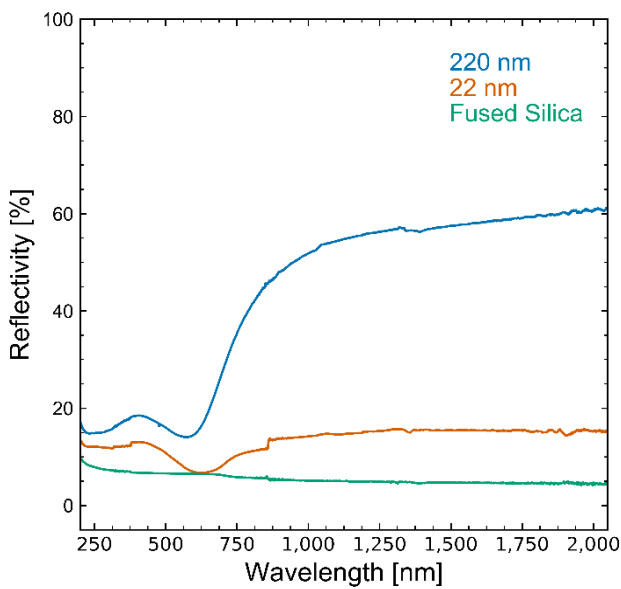

*Fig. 12* Reflectivity measurement in the NIR-VIS-UV range of $RuO_2$ thin films of two different thicknesses and the fused silica substrate. The step at 860 nm is due to a measurement artefact (monochromator switching of the instrument)

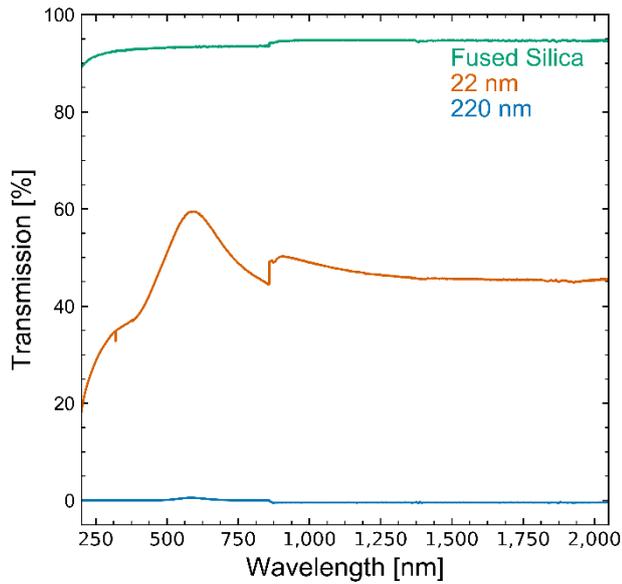

*Fig. 13* Transmission measurement in the NIR-VIS-UV range of RuO2 thin films of two different thicknesses and the fused silica substrate. The step at 860 nm is due to a measurement artefact (monochromator switching of the instrument)

FTIR transmission spectra show that $RuO_2$ blocks wavelengths between 2 and 16 μm effectively, and this effect can be tuned well by decreasing the thickness of the thin film. Using a 23 nm thin film leads to transmission spectra that 'mimicked' the pattern of the specific substrate used. Moreover, the transmission values are 'cut in half' by the ultrathin $RuO_2$ film, which makes it a suitable material to fine tune transmission in the infrared region. The resistivity of a 22 nm thin film, which is 3.4 μΩm, would also make the thin film suitable for applications that require high electrical conductivity.

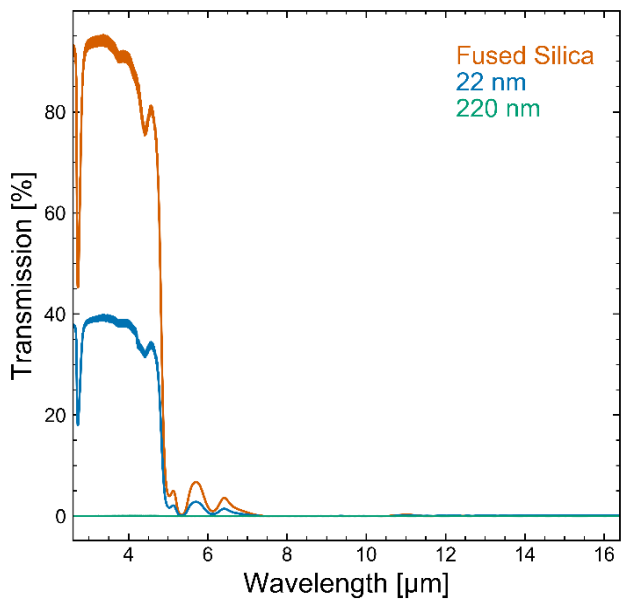

*Fig. 14* Transmission measurement in the IR range of $RuO_2$ thin films of two different thicknesses and the fused silica substrate

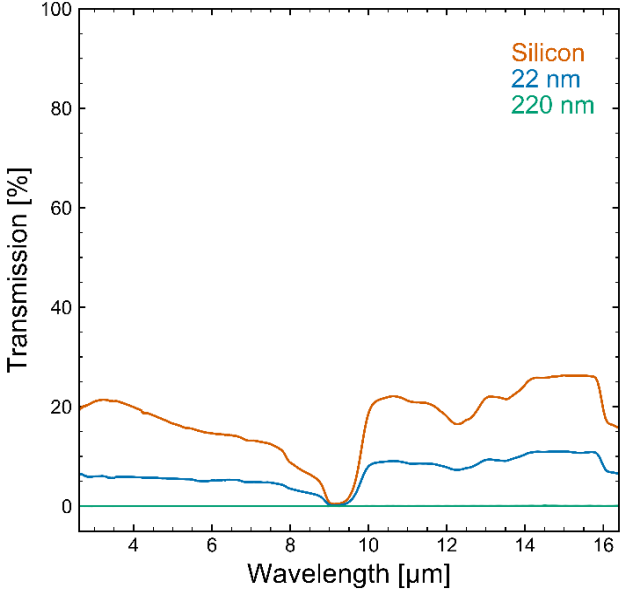

*Fig. 15* Transmission measurement in the IR range of RuO$_2$ thin films of two different thicknesses and the silicon substrate

In the UV-VIS-NIR region, the optical properties of the thin films were investigated in more detail using ellipsometry and the dielectric function ε = ε$_1$+i·ε$_2$ was determined in the range of 300 nm - 1700 nm or 0.73 eV - 4.13 eV, respectively. To model the latter for the evaluation of the ellipsometric spectra, we used a thin-film model that includes a Drude term [33]:

$$\varepsilon(E) = -\frac{AB}{E+iBE}, \qquad (3)$$

where *A* is the amplitude and *B* is the broadening. Moreover, to describe absorption in the NIR we used a Gaussian-broadened oscillator [34], [35]:

$$\varepsilon_2(E) = A e^{-\frac{E-E_0}{\sigma}^2} \text{ with } \sigma = \frac{B}{2}\sqrt{\ln 2}, \qquad (4)$$

where $E_o$ is the center energy. To treat absorption in the VIS-UV region, we used a Cody-Lorentz oscillator [36] (without Urbach absorption, see below):

$$\varepsilon_2(E) = \frac{(E-E_g)^2}{(E-E_g)^2+E_p^2} A E_0 B \frac{E}{(E^2-E_0^2)^2+B^2E^2}, \qquad (5)$$

where $E_o$ is the central energy, $E_g$ = energy gap and $E_p$ defines the energy where the absorption changes from Cody-like to Lorentz-like behavior. This model was originally developed for the description of amorphous semiconductors, but it describes the given polycrystalline layers very well. Furthermore, pole locations and a DC-ε$_1$-offset were used to describe the real part of the dielectric function. Pole locations are given by the following equation:

$$\varepsilon(E) = \frac{A}{E_0^2 - E^2}, \qquad (6)$$

where $E_o$ is the pole outside the measured spectral range. As can be seen in formula (5) above, the Cody-Lorentz model assumes in the region of onset of the absorption above $E_g$ a course of ε$_2$(E) ~ (E - $E_g$)$^2$ and in principle also includes an exponential Urbach absorption term, for which the measurement on the investigated thin films was not sensitive enough. From the analytical expressions of the imaginary part ε$_2$ the corresponding real part ε$_1$ is calculated via the likewise analytical solution of the integral expression:

$$\varepsilon_1(E) = 1 + \frac{2}{\pi} \wp \int_0^\infty \frac{x \cdot \varepsilon_2(x)}{x^2 - E^2} dx ,  \qquad (7)$$

where $\wp$ denotes the principal value of the integral. For the fit based on a Levenberg-Marquardt algorithm, the parameters given above were thus available to fit the measured data. The layer thicknesses were fixed to the values determined from SEM measurements. The measured data could be fitted very well with this model, i.e. with correspondingly low error sums of squares. Table 3 gives the parameters obtained:

*Table 3*: Model parameters obtained from the ellipsometric measurements of 22 nm and 220 nm $RuO_2$ thin films on fused silica for the constitutive elements of the dielectric function

**Pole #1**

| Thickness | Position [eV] | Amplitude |
|---|---|---|
| 22 nm | 6.6166 | 30.786 |
| 220 nm | 4.7808 | 11.568 |

**Pole #2**

| Thickness | Position [eV] | Amplitude |
|---|---|---|
| 22 nm | 0.38696 | 1.4226 |
| 220 nm | 0.36202 | 5.0103 |

**$e_1$-Offset**

| | |
|---|---|
| 22 nm | 1.0728 |
| 220 nm | 2.0362 |

**Drude**

| Thickness | Amplitude | Broadening |
|---|---|---|
| 22 nm | 6.1141 | 2.6591 |
| 220 nm | 8.9504 | 2.75 |

**Gaussian Oscillator**

| Thickness | Amplitude | Center Energy [eV] | Broadening |
|---|---|---|---|
| 22 nm | 729.28 | 0.005418 | 1.3081 |
| 220 nm | 1288.1 | 0.005418 | 1.1235 |

**Cody-Lorentz-Oscillator**

| Thickness | Amplitude | Center Energy [eV] | Broadening | Energy gap [eV] | $E_p$ |
|---|---|---|---|---|---|
| 22 nm | 14.899 | 2.7159 | 3.8408 | 1.8402 | 0.80103 |
| 220 nm | 26.76 | 2.7197 | 3.8512 | 1.7715 | 1.0512 |

Fig. 16 displays the real- and the imaginary part of the thin films dielectric function as a function of the spectral energy $E = \hbar \cdot \omega = h \cdot c / \lambda$ (c = light velocity in vacuum) obtained by using the functions given above.

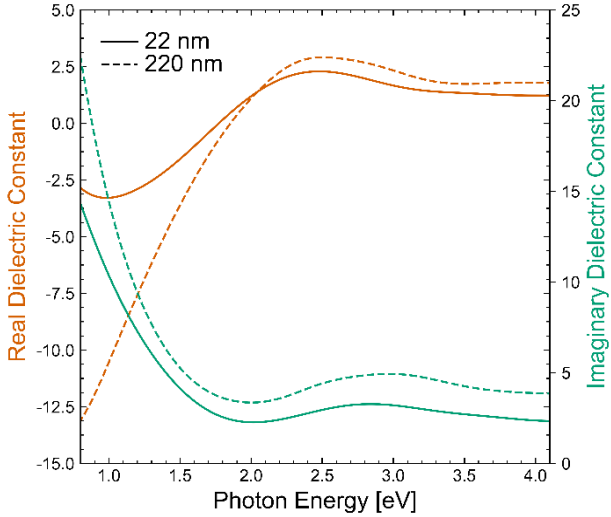

*Fig.* 16 Dielectric constants of a 22 nm and 220 nm thick RuO$_2$ thin film on fused silica calculated from the optical constants derived from ellipsometry measurements.

The dielectric function shows a metal-like behavior, where due to the free charge carrier absorption the imaginary part increases towards lower energies and the real part becomes negative. This behavior is quantitatively weaker than for typical metals, in agreement with the higher electrical resistivity exhibited by the measured thin films compared to metals. Compared to the 220 nm layer, the 22 nm layer shows an average of almost 50% lower $\varepsilon_2$ over the entire spectral range. An interpretation of this phenomenon (e.g. possible increased charge carrier scattering due to microstructural differences), as well as for the significantly stronger drop of $\varepsilon_1$ of the 220 nm layer into the negative at low energies, must be reserved for future investigations. The imaginary part of $\varepsilon(E)$ shows a relative minimum (as a part of metals does) in the range of around 2 eV. Here, comparisons with band structure calculations should offer the possibility to decide whether this can be attributed to a corresponding electronic density of states distribution $Z(E)$.

Also, we found out, that the thin film dielectric functions are almost the same on different substrates (silicon and fused silica), hence similar optical layer properties can be obtained on varying substrates (see Fig. S4 supplementary section).

From the real and imaginary part of the dielectric function, the optical constants refractive index *n* and absorption constant *k* result in:

$$n^2 = \frac{1}{2}\left[\sqrt{\varepsilon_1^2 + \varepsilon_2^2} + \varepsilon_1\right] \quad k^2 = \frac{1}{2}\left[\sqrt{\varepsilon_1^2 + \varepsilon_2^2} - \varepsilon_1\right] \tag{8}$$

Fig. 17 shows the obtained values for the 22 nm and 220 nm thin films on fused silica.

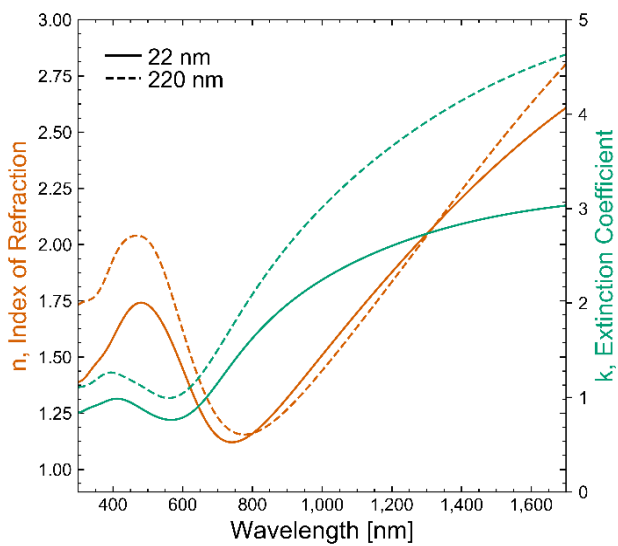

*Fig. 17* Optical constants of a 22 nm and 220 nm thick RuO$_2$ thin film on fused silica derived from ellipsometry measurements.

# 4 Conclusion

RuO$_2$ thin films have been successfully prepared with a novel environmentally-friendly chemical solution deposition process using simply water and acetic acid as solvents. The influence of the annealing temperature on the microstructure and resistivity was investigated and it revealed that dense and smooth thin films with a very low resistivity of 0.89 µΩm can be obtained with an annealing temperature of 600°C. XRD and Raman measurements confirmed that the thin films are phase pure. The electrical characterization showed that the thin films improve in conductivity when the thickness is increased and a metal-like increase in resistivity with increasing temperature. Optical measurements revealed that the thin films are non-transparent, due to their metallic character, but that it is possible to fine tune this behavior by adjusting the thickness. The thermal stability was investigated by post-annealing the samples, and the thin films were stable up to 750°C in synthetic air. However, higher temperature of 850°C led to formation of RuO$_x$ gases with consequent degradation of the film's microstructure.

In conclusion, the high conductivity and thermal, chemical and electrical stability of these simple-to-obtain RuO$_2$ thin films may render them useful as electrodes or buffer layers for a multitude of applications such as ferroelectric and magnetoresistive devices, SOFC, (chemical) sensors, MEMS, geothermal applications, aerospace power electronics and semiconductor devices (e.g. interconnects, memristors, gate contacts). Moreover, the tunable transparency behavior of the thin films makes the material interesting as optical filters for e.g. smart windows and other optoelectronic devices.

**Statements and Declarations**

**Funding** This work was supported by the European Union's Horizon 2020 Research and Innovation Programme under grant agreement Nr. 951774.

**Competing Interests** The authors have no relevant financial or non-financial interests to disclose.

**Author Contributions** Material preparation, data collection and analysis was performed mainly by Martina Angermann; Georg Jakopic collected and analyzed the optical measurements; Thomas Grießer measured and analyzed the XPS spectra; Christine Prietl did the ellipsometry measurements; Klaus Reichmann measured TGA/DSC of the samples. Supervision, funding acquisition and guidance on the paper organization and structure was done by Marco Deluca. The first draft of the manuscript was written by Martina Angermann and all authors commented on previous versions of the manuscript. All authors read, revised and approved the final manuscript.